\documentstyle[12pt]{article}
\font\mybb=msbm10 at 10pt
\def\bb#1{\hbox{\mybb#1}}
\def\Z {\bb{Z}}
\def\R {\bb{R}}

\begin{document}

\begin{flushright}
UG--3/96\\
SU-ITP-19\\
CERN--TH/96--106\\
{\bf hep-th/9605059}\\
July $25$th, $1996$
\end{flushright}

\begin{center}


\baselineskip16pt

{\Large {\bf Stationary Axion/Dilaton Solutions and Supersymmetry}}

\vspace{.3cm}

{\large
{\bf
Eric Bergshoeff\footnote{\tt bergshoe@th.rug.nl}${}^{\dagger}$,
Renata Kallosh\footnote{\tt kallosh@physics.stanford.edu}${}^{\Diamond}$,
Tom\'as Ort\'{\i}n\footnote{\tt tomas@mail.cern.ch}${}^{\ddag}$
}
}
\\
\vspace{.2cm}
${}^{\dagger}$
{\it Institute for Theoretical Physics, University of Groningen}\\
{\it Nijenborgh 4, 9747 AG Groningen, The Netherlands}\\
\vspace{.2cm}
${}^{\Diamond}$
{\it Physics Department, Stanford University, Stanford  CA 94305,
U.S.A.}\\
\vspace{.2cm}
${}^{\ddag}$
{\it C.E.R.N.~Theory Division, CH-1211, Gen\`eve 23, Switzerland}

\vspace{.3cm}


{\bf Abstract}

\end{center}

\begin{quotation}

\small

\baselineskip16pt

We present a new set of supersymmetric stationary solutions of pure
$N=4,d=4$ supergravity (and, hence, of low-energy effective string
theory) that generalize (and include) the Israel-Wilson-Perj\'es
solutions of Einstein-Maxwell theory.  All solutions have $1/4$ of the
supersymmetries unbroken and some have $1/2$.  The full solution is
determined by two arbitrary complex harmonic functions ${\cal
  H}_{1,2}$ which transform as a doublet under $SL(2,\R)$ $S$~duality
and $N$ complex constants $k^{(n)}$ that transform as an $SO(N)$
vector. This set of solutions is, then, manifestly duality invariant.
When the harmonic functions are chosen to have only one pole, all the
general resulting point-like objects have supersymmetric rotating
asymptotically Taub-NUT metrics with $1/2$ or $1/4$ of the
supersymmetries unbroken.  The static, asymptotically flat metrics
describe supersymmetric extreme black holes.  Only those breaking
$3/4$ of the supersymmetries have regular horizons.  The stationary
asymptotically flat metrics do not describe black holes when the
angular momentum does not vanish, even in the case in which $3/4$ of
the supersymmetries are broken.

\end{quotation}

\begin{flushleft}
CERN--TH/96--106\\
\end{flushleft}

\newpage

\pagestyle{plain}


\section*{Introduction}

The theory of pure $N=4,d=4$ supergravity without vector multiplets
presents an interesting and relatively simple model to study various
bosonic solutions with unbroken supersymmetry. It has a richer
structure than pure $N=2,d=4$ supergravity and, in particular, this
theory allows for configurations with either $1/2$ or $1/4$ of the
$N=4$ supersymmetries unbroken, while in $N=2,d=4$ supergravity only
$1/2$ of the $N=2$ supersymmetries can be unbroken \cite{kn:WO}. This
is related to the fact that the central charge of the $N=2$ theory is
replaced in the $N=4$ theory by a central charge {\it matrix}. In the
appropriate basis, the $N=4$ supersymmetry algebra gives rise to two
Bogomol'nyi bounds that can be saturated independently \cite{kn:FSZ}.
When only one is saturated, only $1/4$ of the supersymmetries are
unbroken.  When both are simultaneously saturated $1/2$ of the $N=4$
supersymmetries are unbroken.

Static axion/dilaton black holes with $1/4$ of the supersymmetries
unbroken were first found in Refs.~\cite{kn:KLOPP,kn:O1,kn:KO}.  The
most general family of solutions, presented in Ref.~\cite{kn:KO},
describes an arbitrary number of extreme black holes with (almost)
arbitrary electric and magnetic charges in the six $U(1)$ gauge groups
and non-trivial complex moduli, {\it i.e.}~non-trivial dilaton and
axion fields which are combined in a single complex scalar field.  The
constraint that the charges have to satisfy is a Bogomol'nyi identity
and it is related to the existence of unbroken supersymmetry
\cite{kn:Bo,kn:WO,kn:FSZ,kn:GWG}. The solutions of this family
generically have $1/4$ of the supersymmetries unbroken, but some (in
fact a whole subfamily) have $1/2$ of the supersymmetries unbroken.

Much work has also been devoted to the so-called axion/dilaton gravity
theory which is a truncation of $N=4,d=4$ supergravity with only one
vector field
\cite{kn:KKOT,kn:Ro,kn:GK,kn:GK2,kn:GK3,kn:GGK,kn:GB,kn:KY}.  However,
as we will extensively discuss, the presence of only one vector field
is insufficient to generate all the interesting metrics.  In
particular, in the supersymmetric limit, with only one vector field
one cannot get those with $1/4$ of the supersymmetries unbroken.

Another interesting feature of pure $N=4,d=4$ supergravity is that it
is the simplest model that exhibits both $S$~and $T$~dualities.  The
group $SL(2,\R)$ (quantum-mechanically broken to $SL(2,\Z)$) acts on
the vector fields by interchanging electric and magnetic fields and
acts on the complex scalar by fractional-linear transformations that,
in particular, include the inversion of the complex scalar.  In the
old supergravity days these transformations were not physically
understood.  They just were simply there.  However, in the framework
of string theory the complex scalar has a physical meaning because it
contains the string coupling constant (the dilaton) and it was
conjectured in Refs.~\cite{kn:FILQ,kn:SJR} that this symmetry could be a
non-perturbative symmetry relating the strong and weak-coupling
regimes of string theory: $S$~duality.

The $T$~duality group is $SO(6)$ (again, quantum-mechanically broken to
$SO(6,\Z)$) and rotates amongst them the six Abelian vector fields of
$N=4,d=4$ supergravity. Its physical meaning can only be found in string
theory.

The family of solutions presented in Ref.~\cite{kn:KO} is invariant
under both dualities and, therefore, it contains all the solutions that
can be generated by using them. Taking into account the number of
charges that can be assigned arbitrarily to each black hole and the fact
that a Bogomol'nyi bound has to be saturated in order to have unbroken
supersymmetry it is reasonably to expect that these are the most general
static black hole solutions of this theory with unbroken supersymmetry.

An interesting aspect of the static black hole solutions with unbroken
supersymmetry in pure $N=4$ supergravity is the intriguing relation
that seems to hold between the number of unbroken supersymmetries and
the finiteness of the area of the horizon. All those with only $1/4$
of the supersymmetries unbroken have finite area, while those with
$1/2$ do not have a regular horizon\footnote{This relation no longer
exist when matter is added to the pure supergravity theory or in other
supergravities ($N=8$). An example is provided by theextreme
$a=1/\sqrt{3}$ dilaton black hole, which is a solution of the
low-energy heterotic string theory compactified in a six-torus ($N=4$
supergravity coupled to 22 vector multiplets), which has a singular
horizon and $1/4$ of the $N=4$ supersymmetries unbroken. This Black
hole is also a solution of the $N=8$ supergravity theory with only
$1/8$ of the supersymmetries (i.e.~the same as in $N=4$) unbroken
\cite{kn:KhO}.}.

But the static ones are not the only solutions with unbroken
supersymmetry of $N=4,d=4$ supergravity.  Some stationary solutions are
also known \cite{kn:KKOT}, but all of them have $1/2$ of the
supersymmetries unbroken and none of them represents a black hole with
regular horizon.  It is clear, though, that stationary solutions with
$1/4$ of the supersymmetries unbroken must exist.  In particular, it
must be possible to embed the stationary solutions of pure $N=2,d=4$
supergravity into $N=4,d=4$ supergravity.  These were found in
Ref.~\cite{kn:T} and turned out to be the Israel-Wilson-Perj\'es (IWP)
\cite{kn:IWP} family of metrics.  It is a general feature that solutions
of $N=2,d=4$ supergravity with $1/2$ of the supersymmetries unbroken
have $1/4$ of them unbroken when embedded in $N=4,d=4$.  The best-known
example \cite{kn:GH,kn:KLOPP} is the extreme Reissner-Nordstr\"om (RN)
black hole and the entire Majumdar-Papapetrou (MP) \cite{kn:MP} family
of metrics describing many extreme RN black holes in equilibrium
\cite{kn:HaHa}.  Therefore, one would  expect to find the IWP metrics as
solutions of $N=4,d=4$ supergravity with $1/4$ of the supersymmetries
unbroken, and we will present here the explicit embedding.

The main goal of this paper is to find the most general class of
supersymmetric stationary solutions of $N=4,d=4$ supergravity, which
should include the so-far unknown stationary solutions with $1/4$ of the
supersymmetries unbroken (among them the IWP metrics), those with $1/2$
presented in Ref.~\cite{kn:KKOT} as a particular case and all the static
solutions of Ref.~\cite{kn:KO} in the static limit. We will present and
study this general class of solutions that we will call SWIP because they
generalize the IWP solutions.

We can point out already at this stage that the correspondence between
finiteness of the black-hole area and $1/4$ of the supersymmetries
unbroken will not hold for the stationary solutions, because the IWP
metrics are part of them and they are known to be singular except in the
static limit \cite{kn:HaHa}.  The best-known example of this fact is the
Kerr-Newman metric, which reaches the extreme limit $m-|q|=J$ much
before it reaches the supersymmetric limit $m=|q|$.  In this limit it
has a naked singularity and it is an IWP solution.  Similar results have
been found in the context of the low-energy heterotic string effective
action \cite{kn:CY} which is equivalent to $N=4,d=4$ supergravity
coupled to $22$ vector multiplets.

A further reason to study the simple model of $N=4,d=4$ supergravity
is the fact that any solution of this theory can be embedded into
$N=8,d=4$ supergravity, whose solutions are interesting from the point
of view of $U$~duality \cite{kn:HT}.  By looking for the most general
supersymmetric solutions with $1/4$ of unbroken supersymmetry in $N=4$
theory we may make some progress in the problem of finding the most
general solutions with $1/4$ and $1/8$ of unbroken supersymmetry in
the $N=8$ theory \cite{kn:KhO}.

On the other hand, a reduced version of $N=4$ supergravity with only
two vector fields, offers a nice example of $N=2$ supergravity
interacting with one vector multiplet.  Our $N=4$ solutions will
supply us with solutions of the $N=2$ theory with one vector multiplet
with $1/2$ of unbroken supersymmetry.  Supersymmetric solutions in
$N=2$ supergravity coupled to vector multiplets and hyper-multiplets
are poorly understood.  Only solutions of pure $N=2$ supergravity
\cite{kn:T} as well as the ones reduced from $N=4$ supergravity are
known.  In more general cases, when vector multiplets are included, only
magnetic black hole solutions are known \cite{kn:FKS}.  The analysis
of all supersymmetric stationary solutions of $N=4$ supergravity
performed in this paper will allow us to derive some lessons for the
study of generic electric and magnetic solutions of $N=2$ theory.

It will be particularly useful to reinterpret the results for the
$N=4$ theory in terms of the K\"{a}hler geometry of the $N=2$ theory
\footnote{Static solutions of $N=4$ theory have been identified as
  solutions of $N=2$ supergravity interacting with the vector
  multiplet before \cite{kn:FKS}.}.  We will find that for all of our
new stationary solutions, the metric is described as in \cite{kn:FKS},
in terms of the K\"{a}hler potential $K(X,\overline{X})$.  However, in
addition to that, it depends on a chiral $U(1)$ connection $A_{\mu}$
which breaks hypersurface orthogonality and makes the solutions
stationary.  The appearance of this special geometry object has not
been realized before.  However, since both the K\"{a}hler potential as
well as the $U(1)$ connection are invariant under symplectic
transformations, it is not surprising that both of these functions
show up in the canonical metric of the most general duality-invariant
family of solutions.  This suggests how to find the most general
stationary supersymmetric solutions of $N=2,d=4$ supergravity coupled
to an arbitrary number of vector multiplets. We will comment more on
this in Section~\ref{sec-n2}.

Finally, we would like to remark that point-like objects (black holes
among them) are just one type of the many objects described by the
metrics that we are about to present: by taking two complex harmonic
functions that depend on only two or one spatial coordinate one gets
strings or domain walls, although we will not study them here.

This paper is organized as follows: in Section~\ref{sec-iwp} we describe
the SWIP solutions of pure $N=4,d=4$ supergravity. In
Section~\ref{sec-theory} we present the action and our conventions, in
Section~\ref{sec-solutions} we present the solutions, in
Section~\ref{sec-relation} we describe their relation with previously
known solutions and their behavior under duality transformations is
described in Section~\ref{sec-dual}.  In Section~\ref{sec-n2} we discuss
these solutions from the point of view of stationary supersymmetric
solutions of $N=2$ supergravity coupled to vector multiplets.  In
Section~\ref{sec-search} we study the most general single point-like
object described by a SWIP solution.  In Section~\ref{sec-Taub-NUT} we
study those with NUT charge and no angular momentum (extreme
axion/dilaton Taub-NUT solutions) and in Section~\ref{sec-Rotating} we
study asymptotically flat ({\it i.e.}~zero NUT charge) rotating
solutions showing that there are no supersymmetric rotating black holes
(that is, with regular horizon) with $1/4$ or more supersymmetries
unbroken in $N=4,d=4$ supergravity.  In Section~\ref{sec-susy} we proof
explicitly the supersymmetry of the point-like SWIP metrics and find
their Killing spinors.  Section~\ref{sec-conclusion} contains our
conclusions.


\section{General axion/dilaton IWP solution}
\label{sec-iwp}


\subsection{$N=4,d=4$ supergravity}
\label{sec-theory}

Our conventions are those of Refs.~\cite{kn:KLOPP,kn:KKOT} and are
summarized in an appendix of the first reference, the only difference
being that world indices are underlined instead of carrying a hat.  Our
theory contains a complex scalar $\lambda=a+ie^{-2\phi}$ that
parametrizes an $SL(2,\R)$ coset.  When we consider this theory as part
of a low-energy effective string theory, $a$ is the axion field (the
dual of the usual two-form axion field $B_{\mu\nu}$) and $\phi$ is the
dilaton field.  It also contains the Einstein metric $g_{\mu\nu}$ and
and an arbitrary number $N$ of $U(1)$ vector fields $A_{\mu}^{(n)}$,
$n=1,2,\ldots,N$.

The action is

\begin{eqnarray}
S  & = &
\frac{1}{16\pi}
\int d^{4}x\sqrt{-g}\left\{-R+2(\partial\phi)^{2}
+{\textstyle\frac{1}{2}}e^{4\phi}(\partial a)^{2} \right.
\nonumber \\
& & \\
& &
\left.
-\sum_{n=1}^{N} \left[ e^{-2\phi} F^{(n)}F^{(n)}
+iaF^{(n)}\ {}^{\star}F^{(n)}\right] \right\}\; ,
\nonumber
\end{eqnarray}

When the total number of vector fields is six, this action is identical
to the bosonic part of the action of $N=4,d=4$ supergravity.  When the
total number of vector fields is bigger than six, this action does not
correspond to any supergravity action: additional vector multiplets of
$N=4,d=4$ supergravity would have additional scalars.  We prefer to
leave the number of vector fields arbitrary for the sake of generality.

Sometimes it is convenient to use alternative ways of writing this
action:

\begin{eqnarray}
S  & = &
\frac{1}{16\pi}
\int d^{4}x\sqrt{-g}\left\{-R
+{\textstyle\frac{1}{2}}
\frac{\partial_{\mu}\lambda
\partial^{\mu}\overline{\lambda}}{(\Im{\rm m}\lambda)^{2}}
-i\sum_{n=1}^{N}F^{(n)}\ {}^{\star}\tilde{F}^{(n)}\right\}
\\
& &  \nonumber \\
& &  \nonumber \\
& = &
\frac{1}{16\pi}
\int d^{4}x\sqrt{-g}\left\{-R
+{\textstyle\frac{1}{2}}
\frac{\partial_{\mu}\lambda
\partial^{\mu}\overline{\lambda}}{(\Im{\rm m}\lambda)^{2}}
+2\Re {\rm e}\left( i\lambda\sum_{n=1}^{N}F^{(n)+}F^{(n)+} \right)
\right\}\, ,
\end{eqnarray}

\noindent where we have defined the $SL(2,\R)$-duals\footnote{The
space-time duals are ${}^{\star}F^{(n)\mu\nu}=\frac{1}{2\sqrt{-g}}
\epsilon^{\mu\nu\rho\sigma}F_{\rho\sigma}$, with
$\epsilon^{0123}=\epsilon_{0123}=+i $ .} to the fields
$F_{\mu\nu}^{(n)}= \partial_{\mu} A_{\nu}^{(n)}- \partial_{\nu}
A_{\mu}^{(n)}$

\begin{equation}
\tilde{F}^{(n)}=e^{-2\phi}\,{}^{\star}F^{(n)}-iaF^{(n)}\; ,
\end{equation}

The advantage of using $\tilde{F}^{(n)}$ is that the equations of
motion for the vector fields can be written in this way

\begin{equation}
d{}^{\star}\tilde{F}^{(n)}=0\, ,
\end{equation}

\noindent and imply the local existence of $N$ real vector potentials
$\tilde{A}^{(n)}$ such that

\begin{equation}
\tilde{F}^{(n)}_{\mu\nu}=i\, (\partial_{\mu}\tilde{A}^{(n)}_{\nu}
-\partial_{\nu}\tilde{A}^{(n)}_{\mu})\, .
\end{equation}

The analogous equations $F^{(n)}_{\mu\nu} =\partial_{\mu}A^{(n)}_{\nu}
-\partial_{\nu}A^{(n)}_{\mu}$, are a consequence of the Bianchi
identities

\begin{equation}
d{}^{\star}F^{(n)}=0\, ,
\end{equation}

\noindent (or, obviously, the definition of $F^{(n)}_{\mu\nu}$).

If the time-like components $A_{t}^{(n)}$ play the role of electrostatic
potentials, then the $\tilde{A}^{(n)}_{t}$'s will play the role of
magnetostatic potentials.  A virtue of this formalism is that the
duality rotations can be written in terms of the vector fields $A^{(n)}$
and $\tilde{A}^{(n)}$ instead of the field strengths\footnote{Of course,
this is only valid on-shell, where the $\tilde{A}^{(n)}$s exist, but
$SL(2,\R)$ is only a symmetry of the equations of motion anyway.}:

\begin{eqnarray}
A^{(n)\prime}(x)         & = &
\hskip 0.4 cm \delta A^{(n)}(x) -\gamma \tilde{A}^{(n)}(x)\; ,
\nonumber \\
& & \\
\label{eq:trans}
\tilde{A}^{(n)\prime}(x) & = &
-\beta A^{(n)}(x) +\alpha \tilde{A}^{(n)}(x)\, , \nonumber
\end{eqnarray}

\noindent where $\alpha, \beta, \gamma$ and $\delta$
are the elements of an $SL(2,\R)$ matrix

\begin{equation}
R=
\left(
\begin{array}{cc}
\alpha & \beta  \\
\gamma & \delta \\
\end{array}
\right)\; .
\end{equation}

Note that, since the $\tilde{A}^{(n)}$'s are not independent fields, the
consistency of Eqs.~(\ref{eq:trans}) implies the usual transformation
law of $\lambda$:

\begin{equation}
\lambda^{\prime}(x)=\frac{\alpha\lambda(x)+
\beta}{\gamma\lambda(x)+\delta}\; .
\end{equation}

With no dilaton nor axion ($\lambda=i$), our theory coincides with
the Einstein-Maxwell theory.  In this case $\tilde{F}={}^{\star}F$ and the
consistency of Eqs.~(\ref{eq:trans}) would imply that $R$ is an $SO(2)$
matrix, the duality group being just $U(1)$.


\subsection{The SWIP solutions}
\label{sec-solutions}

Now let us describe the solutions. All the functions entering in the
different fields can ultimately be expressed in terms of two completely
arbitrary complex harmonic functions ${\cal H}_{1}(\vec{x})$ and ${\cal
H}_{2}(\vec{x})$

\begin{equation}
\partial_{\underline{i}}\partial_{\underline{i}} {\cal H}_{1}=
\partial_{\underline{i}}\partial_{\underline{i}} {\cal H}_{2}= 0\, ,
\end{equation}

\noindent and a set of complex constants $k^{(n)}$ that satisfy the
constraints

\begin{equation}
\sum_{n=1}^{N} (k^{(n)})^{2}=0\, ,
\hspace{1cm}
\sum_{n=1}^{N} |k^{(n)}|^{2}={\textstyle\frac{1}{2}}\, ,
\end{equation}

\noindent in the general case\footnote{If ${\cal H}_{1}$ or ${\cal
H}_{2}$ are constant then only the second constraint is necessary.}.
This means that in general we must have at least two non-trivial vector
fields.

The harmonic functions enter through the following two combinations into
the metric\footnote{Here $\epsilon_{123}=+1$\ .}:

\begin{eqnarray}
e^{-2U} & = & 2\ \Im{\rm m}\ ({\cal H}_{1}\overline{\cal H}_{2})\, ,\\
& & \nonumber \\
\partial_{[\underline{i}}\
\omega_{\underline{j}]} & = &
\epsilon_{ijk}\
\Re{\rm e}\
\left( {\cal H}_{1} \partial_{\underline{k}}\overline{\cal H}_{2}
-\overline{\cal H}_{2} \partial_{\underline{k}}{\cal H}_{1} \right)\, .
\end{eqnarray}

The fields themselves are

\begin{eqnarray}
ds^{2}  & = &
e^{2U} (dt^{2}+\omega_{\underline{i}}dx^{\underline{i}})^{2}
-e^{-2U}d\vec{x}^{2}\, ,
\\
& & \nonumber\\
\lambda & = &
\frac{{\cal H}_{1}}{{\cal H}_{2}}\, ,
\\
& & \nonumber \\
A_{t}^{(n)} & = &
2e^{2U}\Re {\rm e}\left( k^{(n)}{\cal H}_{2} \right)\, ,
\\
& & \nonumber \\
\tilde{A}^{(n)}_{t} & = &
-2e^{2U}\Re {\rm e} \left( k^{(n)}{\cal H}_{1} \right)\, .
\end{eqnarray}

Given the time-components $A^{(n)}_{t},\tilde{A}^{(n)}_{t}$'s, all the
components of the ``true'' vector fields $A^{(n)}_{\mu}$ are completely
determined.


\subsection{Relation with previously known solutions}
\label{sec-relation}

First we observe that these solutions reduce to the usual IWP metrics
Refs.~\cite{kn:IWP} when

\begin{equation}
{\cal H}_{1}=i{\cal H}_{2}={\textstyle\frac{1}{\sqrt{2}}} V^{-1}\, ,
\end{equation}

\noindent (in the notation of Ref.~\cite{kn:T}), where $V^{-1}$ is a
complex harmonic function, but now embedded in $N=4$ supergravity.  All
these metrics admit Killing spinors when embedded in $N=2$ supergravity
\cite{kn:T} and generically have $1/2$ of the supersymmetries of $N=2$
supergravity unbroken \cite{kn:KO2}.  In fact, apart from $pp$-waves,
all supersymmetric solutions of $N=2,d=4$ supergravity have IWP metrics
\cite{kn:T}.

However, the only black-hole-type solutions ({\it i.e.}~describing
point-like objects with regular horizons covering the singularities) in
this class are the MP solutions Ref.~\cite{kn:MP}
($V=\overline{V}$ with the right asymptotics) that describe an arbitrary
number of extreme RN black holes in static equilibrium
\cite{kn:HaHa}.  Any amount of angular momentum added to the MP solutions
produces naked singularities.  One can also add NUT charge (general IWP
metrics have angular momentum and NUT charge ), but, then, the spaces
are not asymptotically flat and do not admit a black hole
interpretation.  There are no supersymmetric rotating black holes in
pure $N=2,d=4$ supergravity and, in consequence, in the IWP subclass of
the SWIP solutions of $N=4,d=4$ supergravity there will be no
supersymmetric rotating black holes, either. We will discuss later how
many $N=4,d=4$ unbroken supersymmetries they have.

If we now take

\begin{equation}
{\cal H}_{1}=\frac{i}{\sqrt{2}}\lambda\, ,
\hspace{1.5cm}
{\cal H}_{2}=\frac{i}{\sqrt{2}}\, ,
\end{equation}

\noindent we recover the electric-type axion-dilaton IWP solutions of
Ref.~\cite{kn:KKOT}.  All these solutions have $1/2$ of all $N=4$
supersymmetries unbroken.  Again, all solutions with angular momentum in
this class have naked singularities.  Observe that this class does not
contain any $N=2$ IWP metric and is intrinsically $N=4$.

Finally, if we impose the constraint

\begin{equation}
\partial_{[\underline{i}}\ \omega_{\underline{j}]}=0\, ,
\Rightarrow
\Re{\rm e}\
\left( {\cal H}_{1} \partial_{\underline{k}}\overline{\cal H}_{2}
-\overline{\cal H}_{2} \partial_{\underline{k}}{\cal H}_{1}
\right)=0\, ,
\end{equation}

\noindent which gives static metrics, we recover the solutions of
Ref.~\cite{kn:KO}.  The existence of a constraint on the harmonic
functions in those solutions is implicit in the form in which the two
harmonic functions were given.  However, when those solutions were found
by a lengthy and not transparent process of covariantization with
respect to $SL(2,\R)$ duality rotations (first discussed in this context
in Ref.~\cite{kn:STW}) it was almost impossible to see that the
relations between the two complex harmonic functions could be expressed
as just the effect of imposing that the metrics are static, as opposite
to stationary.

The black-hole solutions in this class have either $1/2$ or $1/4$ of the
supersymmetries unbroken \cite{kn:KLOPP,kn:O2}.  The complexity of the
constraints between the complex harmonic functions made a proof of the
supersymmetry of the general solution very difficult to obtain.  This
class trivially contains all the static $N=2$ IWP metrics, and also the
usual extreme dilaton black holes \cite{kn:G,kn:GM,kn:GHS} from which
they were obtained by generalization \cite{kn:O1} and
$SL(2,\R)$-covariantization \cite{kn:KO}.

Finally, we should mention the relation with solutions of the
low-energy heterotic string theory compactified on a six-torus
\cite{kn:MS}.  From the supergravity point of view, this theory is
nothing but $N=4,d=4$ supergravity coupled to $22$ vector
supermultiplets.  Our solution is related to truncations in which all
$22$ matter vector fields (and scalars) vanish.  This truncation
corresponds to the case in which the six vector fields that come from
the ten-dimensional axion are identified (up to a convention-dependent
sign) with the six vector fields that come from the ten-dimensional
metric and the remaining $16$ vector fields vanish.  The most general
static solution of the low-energy heterotic string effective action
compactified on a six-torus and {\it given in terms of independent
  harmonic functions} was recently found in Ref.~\cite{kn:CT} and
rediscovered in Ref.~\cite{kn:Ra}.  The truncation typically reduces
the number of independent harmonic functions by a half.  Then, some of
the static solutions in the SWIP class, with only two real independent
harmonic functions correspond to some of the solutions in
Ref.~\cite{kn:CT} with two independent harmonic functions.  It is
difficult to make more accurate comparisons between these two families
of metrics because they are given in two very different settings.

Another stationary solution of the low-energy heterotic string effective
action depending on just one real harmonic function $F^{-1}$ was
recently constructed in Ref.~\cite{kn:HS}.  This solution breaks $1/2$
of the supersymmetries and, when the matter vector fields are set to
zero, they correspond to one of the $N=4$ solutions in
Ref.~\cite{kn:KKOT}.

We will say more about this and other solutions in
Section~\ref{sec-search} where we will discuss the relation between
other known solutions and the SWIP metrics in the single point-like
object (spherically or axially symmetric) case.


\subsection{Duality properties}
\label{sec-dual}

Since our general SWIP solution is a straightforward generalization of
the solutions in Ref.~\cite{kn:KO}, it shares many of their properties,
in particular those concerning its behavior under $SL(2,\R)$
transformations: the whole family transforms into itself under
$SL(2,\R)$. Individual solutions in this family are simply interchanged
by $SL(2,\R)$ transformations, and, therefore, the effect of the
transformations can be expressed by substituting ${\cal H}_{1,2}$ and
the $k^{(n)}$s by its {\it primed} counterparts: ${\cal
H}_{1,2}^{\prime}$ and $k^{(n)\prime}$.

So, what is the action of $SL(2,\R)$ on them? ${\cal H}_{1,2}$ transform
as a doublet (that is, linearly) under $SL(2,\R)$ while the $k^{(n)}$'s
are invariant.  To be more precise, observe first that ${\cal
H}_{1},{\cal H}_{2}$ and the $k^{(n)}$'s are defined up to a complex
phase: if we multiply ${\cal H}_{1}$ and ${\cal H}_{2}$ by the same
complex phase and the constants $k^{(n)}$ by the opposite phase, the
solution is invariant.  An $SL(2,\R)$ rotation as we have defined it
transforms ${\cal H}_{1,2}$ as a doublet up to a complex phase and
scales the $k^{(n)}$'s by the opposite phase but we can always absorb
that phase as we just explained.

Incidentally, the solution is also $SO(N)$ covariant: the constants
$k^{(n)}$ (and, hence, the $N$ $U(1)$ vector fields) transform as an
$SO(N)$ vector\footnote{$SO(6,\Z)$ is the duality group of the pure
$N=4$ truncation of heterotic string theory compactified on a
six-torus.}.

Therefore, this solution has all the duality symmetries of $N=4,d=4$
supergravity ``built in'' and nothing more general can be generated by
duality rotations. In a sense they are the ``true'' generalization of
the IWP solutions which also have built in the duality symmetries of
$N=2,d=4$ supergravity: a $U(1)$ electric-magnetic duality rotation of
the single vector field of $N=2,d=4$ supergravity (the only duality
symmetry of this theory) corresponds to multiplying the complex
harmonic function $V^{-1}$ by a complex phase, giving another IWP
solution with the same metric etc.

On the other hand, all these metrics have a time-like isometry and one
can perform further $T$~duality (Buscher \cite{kn:B}) transformations in
the time direction. We will not attempt to study here the result of
these transformations.


\subsection{Relation with general $N=2$ solutions}
\label{sec-n2}

In Ref.~\cite{kn:FKS}, whose conventions we follow in this Section,
some static extremal magnetic black hole solutions of $N=2,d=4$
supergravity coupled to an arbitrary number of vector multiplets were
constructed in the case in which the ratios of the complex
$X^{\Lambda}$ are real. The metric turns out to depend exclusively on
the K\"{a}hler potential $K(Z,\overline{Z})$, where $Z^{\Lambda}
=X^{\Lambda}/X^{0}$:

\begin{equation}
ds^{2} = e^{2U}dt^{2} -e^{-2U}d\vec{x}^{2}\, ,
\hspace{1cm}
e^{2U}= e^{K(Z,\overline{Z})-K_{\infty}}\, .
\end{equation}

It was then realized that the solutions presented in
Ref.~\cite{kn:KO} could also be described in terms of the K\"{a}hler
potential associated to pure $N=4,d=4$ supergravity with only two
vector fields interpreted as $N=2,d=4$ supergravity coupled to a single
vector multiplet. For this theory, the prepotential is $F(X)=2X^{0}X^{1}$.
Taking $X^{0}$ and $X^{1}$ to be two suitably normalized complex harmonic
functions

\begin{equation}
X^{0}=i {\cal H}_{2}\, ,
\hspace{1cm}
X^{1}= {\cal H}_{1}\, ,
\end{equation}

\noindent one immediately arrives to the  K\"{a}hler potential

\begin{eqnarray}
e^{-K(X,\overline{X})} & = &
\overline{X}^{\Lambda} N_{\Lambda\Sigma} X^{\Sigma}\nonumber \\
& & \nonumber \\
& = & 2 \Im {\rm m}\ ( {\cal H}_{1}\overline{\cal H}_{2}) \nonumber \\
& & \nonumber \\
& = & e^{-2U} \, .
\end{eqnarray}

Therefore the K\"{a}hler potential provides the factor $e^{2U}$ in the
metric. With the complex harmonic functions ${\cal H}_{1,2}$
constrained as in Ref.~\cite{kn:KO}, this is all there is to it.
However, if the harmonic functions ${\cal H}_{1,2}$ are not
constrained, there is another geometrical object which is a symplectic
invariant and that in this case does not vanish: the chiral connection

\begin{eqnarray}
A_{\mu} & = & {\textstyle\frac{i}{2}} N_{\Lambda\Sigma}\
\left[ \overline{X}^{\Lambda} \partial_{\mu} X^{\Sigma}
-(\partial_{\mu}\overline{X}^{\Lambda}) X^{\Sigma}\right] \nonumber \\
& & \nonumber \\
& = &  -2 \Re {\rm e}\ ({\cal H}_{1}\partial_{\mu}\overline{\cal H}_{2}
-\overline{\cal H}_{2} \partial_{\mu} {\cal H}_{1}) \nonumber \\
& & \nonumber \\
& = & -\delta_{\mu}{}^{k}\ \epsilon_{ijk}\
\partial_{[\underline{i}} \omega_{\underline{j}]}\, .
\end{eqnarray}

Thus, the chiral connection naturally gives us the three-vector
$\omega_{\underline{i}}$ which codifies the information about angular
momentum, NUT charge etc. Imposing the condition that the metric is
static can be simply expressed as the vanishing of the chiral
connection:

\begin{equation}
A_{\mu}=0\, .
\end{equation}

It is surprising to some extent and highly nontrivial that this
constraint does not constrain the values of the electric and magnetic
charges and the masses and that they are essentially arbitrary (as
long as the Bogomol'nyi bound is saturated). This constraint seems to
have a purely (space-time) geometrical content.

We see that all the elements appearing in the general metric of the
SWIP solution have a special geometrical meaning.  Hence, it is only
natural to try to extend this scheme to more general $N=2,d=4$ theories
with an arbitrary number of vector multiplets and arbitrary
prepotential $F(X)$. We conjecture that the metric of most general
supersymmetric stationary solution of $N=2,d=4$ supergravity coupled to vector
multiplets can be written in the form:

\begin{equation}
ds^{2} = e^{K(X,\overline{X})} (dt^{2}
+\omega_{\underline{i}}dx^{\underline{i}})^{2}
-e^{-K(X,\overline{X})}d\vec{x}^{2}\, ,
\end{equation}

\noindent where $K(X,\overline{X})$ is the K\"{a}hler potential
associated to the corresponding holomorphic prepotential $F(X)$ and
$\omega_{\underline{i}}$ is determined by the chiral connection
$A_{\mu}$ as before. A natural step that would generalize the results
obtained in Ref.~\cite{kn:FKS} for static purely magnetic solution
would be to obtain static dyonic solutions by using the constraint of
the vanishing of the chiral connection.


\section{Point-like SWIP solutions}
\label{sec-search}

The most general choice of complex harmonic functions for a point-like
object is

\begin{equation}
{\cal H}_{1}  =  \chi_{0} + \frac{\chi_{1}}{r_{1}}\, ,
\hspace{1.5cm}
{\cal H}_{2} = \psi_{0} + \frac{\psi_{1}}{r_{2}}\, ,
\end{equation}

\noindent where $\chi_{0},\chi_{1},\psi_{0},\psi_{1}$ are arbitrary
complex constants and $r_{1,2}^{2} =(\vec{x}-\vec{x}_{1,2}) \cdot
(\vec{x}-\vec{x}_{1,2})$ where the constants $\vec{x}_{1,2}$ are
arbitrary and complex.  Up to shifts in the coordinate $z$, the most
general possibility compatible with having a single point-like object is
$r_{1}^{2} =r_{2}^{2} =x^{2}+y^{2}+(z-i\alpha)^{2}$.

We have, then, at our disposal, $9+2n$ real integration constants,
including the $k^{(n)}$'s.  After imposing the normalization of the
metric at infinity and the three real constraints that the $k^{(n)}$'s
have to satisfy in order to have a solution, it seems that we are left
with $5+2n$ independent integration constants.  However, if one
multiplies both ${\cal H}_{1,2}$ by the same complex phase, the solution
remains invariant, and so we only have $4+2n$ meaningful independent
integration constants at our disposal with $n\geq 2$ in the generic
case.

On the other hand, the maximum number of independent physical parameters
that we can have seems to be\footnote{Although some sort of ``no-hair''
theorem probably holds for this theory, none has yet been proven.}
$5+2n$: the mass $m$, the NUT charge $l$, the angular momentum $J$, the
complex moduli $\lambda_{0}$ (which is the value of the complex scalar
at infinity) and $2n$ electric and magnetic charges $(q^{(n)},p^{(n)})$.
However, when the field configurations have unbroken supersymmetry,
there is at least one constraint between them: the Bogomol'nyi identity.
Therefore we only expect $4+2n$ independent physical parameters in the
supersymmetric case.  We are going to show that all these solutions
satisfy the Bogomol'nyi identity and, since the number of independent
physical parameters matches the number of integration constants, we
expect them to be the most general axisymmetric solutions of our theory
with (at least) $1/4$ of the supersymmetries unbroken.

Studying the asymptotic behavior of the different fields (see later) we
find that the integration constants are related to the physical
parameters of the solution\footnote{Again we use the same definitions
for electric and magnetic charges as in Ref.~\cite{kn:KO}.  The only
difference is that the mass $m$ has to be substituted by the complex
combination ${\cal M}=m+il$, where $l$ is the NUT charge, in the
definition of the complex scalar charge
$\Upsilon=-2\sum_{n}\overline{\Gamma^{(n)}}^{2}/{\cal M}$.} by

\begin{equation}
\begin{array}{rclrcl}
\chi_{0} & = & \frac{1}{\sqrt{2}} e^{\phi_{0}}\lambda_{0} e^{i\beta}\, ,
&
\chi_{1} & = & \frac{1}{\sqrt{2}} e^{\phi_{0}} e^{i\beta} (\lambda_{0}
{\cal M} +\overline{\lambda}_{0}\Upsilon)\, ,\\
& & & & & \\
\psi_{0} & = & \frac{1}{\sqrt{2}} e^{\phi_{0}} e^{i\beta}\, ,
&
\psi_{1} & = & \frac{1}{\sqrt{2}} e^{\phi_{0}} e^{i\beta}({\cal M} +
\Upsilon)\, ,\\
& & & & & \\
k^{(n)} & = & - e^{-i\beta}
\begin{tabular}{c}
${\cal M} \Gamma^{(n)} +\overline{\Upsilon\Gamma^{(n)}}$ \\
\hline
$|{\cal M}|^{2}-|\Upsilon|^{2}$ \\
\end{tabular}
\, ,
&
\alpha & = & J/m\, ,\\
\end{array}
\end{equation}

\noindent where $J$ is the angular momentum and $\beta$ is an arbitrary
real number which does not play any physical role (but transforms under
$S$~duality according to our prescription).

Observe that these identifications have been made under the assumption
$m\neq 0$. There are also massless solutions in this class.

The functions ${\cal H}_{1},{\cal H}_{2},e^{-2U}$ and
$\omega_{\underline{i}}$ take the form

\begin{eqnarray}
{\cal H}_{1} & = & {\textstyle\frac{1}{\sqrt{2}}} e^{\phi_{0}}
\lambda_{0} e^{i\beta} \left(\lambda_{0} + \frac{\lambda_{0}{\cal M}
+\overline{\lambda}_{0}\Upsilon}{r} \right)\, ,
\label{eq:h1point}
\\
& & \nonumber \\
{\cal H}_{2} & = & {\textstyle\frac{1}{\sqrt{2}}} e^{\phi_{0}}
e^{i\beta} \left(1 + \frac{{\cal M} +\Upsilon}{r} \right)\, , \\
& & \nonumber \\
e^{-2U} & = & 1 + 2\ \Re {\rm e}\left(\frac{\cal M}{r}\right)
+\frac{(|{\cal M}|^{2}-|\Upsilon|^{2})}{|r|^{2}}\, ,
\label{eq:h2point}
\\
& & \nonumber \\
\partial_{[\underline{i}}\ \omega_{\underline{j}]} & = &
\epsilon_{ijk}\
\Im{\rm m}\
\left[{\cal M}\ \partial_{\underline{k}}\frac{1}{r}
+(|{\cal M}|^{2}-|\Upsilon|^{2})
\frac{1}{r}\partial_{\underline{k}}\frac{1}{\overline{r}}\right]\, .
\end{eqnarray}

That the above identifications between physical parameters and
integration constants are correct becomes evident when we switch
from Cartesian to oblate spheroidal coordinates $(\rho,\theta,\varphi)$

\begin{equation}
\left\{
\begin{array}{rcl}
x\pm iy & = & \sqrt{\rho^{2}+\alpha^{2}}\ \sin{\theta}\ e^{\pm
i\varphi}\, , \\
& & \\
z & = & \rho\cos{\theta}\, ,\\
& & \\
d\vec{x}^{2} & = &
(\rho^{2}+\alpha^{2}\cos^{2}\theta) (\rho^{2}+\alpha^{2})^{-1}d\rho^{2}
+(\rho^{2} +\alpha^{2}\cos^{2}\theta)\ d\theta^{2}\\
& & \\
& &
+(\rho^{2}+\alpha^{2})\sin^{2}\theta\ d\varphi^{2}\, ,\\
\end{array}
\right.
\end{equation}

\noindent in which only the component $\omega_{\varphi}$ does not vanish
and takes the form\footnote{Observe that $\rho$ can take positive or
negative values since its sign is not determined by the coordinate
transformation. We have used $r=\rho +i\alpha\cos\theta$.}

\begin{eqnarray}
\omega_{\varphi} & = & 2\cos\theta\ l
+\alpha \sin^{2} \theta\ (e^{-2U} -1) \nonumber \\
& & \\
& = & 2\cos\theta\ l +\alpha \frac{\sin^{2}\theta}{\rho^{2}
+\alpha^{2}\cos^{2}\theta} \left[ 2m\rho + 2l\alpha \cos\theta
+ (|{\cal M}|^{2} -|\Upsilon|^{2})\right]\, , \nonumber
\end{eqnarray}

\noindent while $e^{-2U}$ takes the form

\begin{equation}
e^{-2U}= \frac{\rho^{2} +\alpha^{2}\cos^{2}\theta +2m\rho +2l\alpha
\cos\theta +(|{\cal M}|^{2} -|\Upsilon|^{2})}{\rho^{2}
+\alpha^{2}\cos^{2}\theta}\, .
\end{equation}

The full general metric can be written in the standard way after a shift
in the radial coordinate: $\hat{\rho}=\rho +m +|\Upsilon|$

\begin{equation}
ds^{2} = \frac{\Delta -\alpha^{2}\sin^{2}\theta}{\Sigma}
\left( dt -\omega d\varphi \right)^{2} -
\Sigma \left(\frac{d\hat{\rho}^{2}}{\Delta} +d\theta^{2} +
\frac{\Delta\sin^{2}\theta d\varphi^{2}}{\Delta
-\alpha^{2}\sin^{2}\theta}\right)\, ,
\label{eq:ptmet1}
\end{equation}

\noindent where

\begin{eqnarray}
\omega & = & -\omega_{\varphi} = \frac{2}{\Delta
-\alpha^{2}\sin^{2}\theta}
\left\{l \Delta \cos\theta +\alpha\sin^{2}\theta
\left[m (\hat{\rho} -(m + |\Upsilon|) ) \right. \right. \nonumber \\
& & \nonumber \\
& &
\left.
\left.
+{\textstyle\frac{1}{2}}(|{\cal
M}|^{2} -|\Upsilon|^{2}) \right] \right\}\, ,
\label{eq:ptmet2}
\\
& & \nonumber \\
\Delta & = & \hat{\rho}[\hat{\rho} -2(m+|\Upsilon|)] +\alpha^{2}
+(m+|\Upsilon|)^{2}\, ,
\label{eq:ptmet3}
\\
& & \nonumber \\
\Sigma & = & \hat{\rho} (\hat{\rho} -2 |\Upsilon|)
+(\alpha\cos\theta +l)^{2}\, .
\label{eq:ptmet4}
\end{eqnarray}

Before comparing this metric with other rotating Taub-NUT solutions in
the literature, we make the following observation: by construction, the
following relation between the physical parameters is always obeyed in
this class of solutions:

\begin{equation}
|{\cal M}|^{2} + |\Upsilon|^{2}-4\sum_{n}|\Gamma^{(n)}|^{2}=0\, .
\end{equation}

This is the usual expression of the Bogomol'nyi identity in pure $N=4$
supergravity \cite{kn:O1}, and it is valid for solutions with $1/2$ or
$1/4$ of the supersymmetries unbroken. Multiplying by $|{\cal M}|^{2}$
and using the expression of $\Upsilon$ in terms of ${\cal M}$ and the
$\Gamma^{(n)}$'s we observe that we can always rewrite it in this way:

\begin{equation}
(|{\cal M}|^{2}-|{\cal Z}_{1}|^{2})
(|{\cal M}|^{2}-|{\cal Z}_{2}|^{2})=0\, ,
\end{equation}

\noindent where $|{\cal Z}_{1}|$ and $|{\cal Z}_{2}|$ can be identified
with the two different skew eigenvalues of the central charge matrix
\cite{kn:WO,kn:FSZ}.  The above identity indicated that one of the two
possible Bogomol'nyi bounds

\begin{equation}
|{\cal M}|^{2}\geq |{\cal Z}_{1,2}|^{2}\, ,
\end{equation}

\noindent is always saturated and, therefore, $1/4$ of the
supersymmetries of pure $N=4$ supergravity are always unbroken.

For only two vector fields, the central charge eigenvalues are linear in
electric and magnetic charges:

\begin{eqnarray}
{\cal Z}_{1} & = & \sqrt{2}(\Gamma^{(1)} +i \Gamma^{(2)})\, ,
\nonumber \\
& & \\
{\cal Z}_{2} & = & \sqrt{2}(\Gamma^{(1)} -i \Gamma^{(2)})\, ,
\end{eqnarray}

\noindent but, in general, we have the non-linear expression\footnote{We
stress that only when there are six or less vector fields ${\cal
Z}_{1,2}$ have an interpretation in terms of pure $N=4,d=4$
supergravity.}

\begin{equation}
{\textstyle\frac{1}{2}}|{\cal Z}_{1,2}|^{2} =
\sum_{n}|\Gamma^{(n)}|^{2}
\pm \left[\left(\sum_{n} |\Gamma^{(n)}|^{2}\right)^{2}
-|\sum_{n}\Gamma^{(n)2}|^{2}\right]^{\frac{1}{2}}\, .
\end{equation}

When both central charge eigenvalues are equal $|{\cal Z}_{1}|
=|{\cal Z}_{2}|$ and, therefore, $1/2$ of the supersymmetries are
unbroken, it is easy to prove that

\begin{equation}
|{\cal M}|^{2} =|\Upsilon|^{2} =|{\cal Z}_{1,2}|^{2} =
2|\sum_{n}\Gamma^{(n)2}|\, .
\end{equation}

Now we are ready to compare the metric
Eqs.~(\ref{eq:ptmet1})-(\ref{eq:ptmet4}) with other general rotating
asymptotically Taub-NUT metrics of $N=4,d=4$ supergravity solutions.
The most general metric of this kind was given in Eqs.~(31)-(35) of
Ref.~\cite{kn:GK} and has the same general form Eq.~(\ref{eq:ptmet1})
but now with

\begin{eqnarray}
\omega & = & \frac{2}{\Delta -\alpha^{2}\sin^{2}\theta}
\left\{l \Delta \cos\theta +\alpha\sin^{2}\theta
\left[m (r-r_{-})+l(l-l_{-}) \right] \right\}
\label{eq:ptmetG2}
\\
& & \nonumber \\
\Delta & = & (r-r_{-})(r-2m) +\alpha^{2} +(l-l_{-})^{2}\, ,
\label{eq:ptmetG3}
\\
& & \nonumber \\
\Sigma & = & r(r -r_{-}) +(\alpha\cos\theta +l)^{2} -l_{-}^{2}\, ,
\label{eq:ptmetG4}
\end{eqnarray}

\noindent where the constants $r_{-}, l_{-}$ are related to the electric
charge ${\cal Q}$, the mass and Taub-NUT charge by

\begin{equation}
r_{-}  = \frac{m |{\cal Q}|^{2}}{|{\cal M}|^{2}}\, ,
\hspace{1.5cm}
l_{-}  =  \frac{l |{\cal Q}|^{2}}{2|{\cal M}|^{2}}\, .
\end{equation}

First, note that this metric is not supersymmetric in general.  It
becomes supersymmetric when $|{\cal M}|=\sqrt{2} |{\cal Q}|$.  It is
now very easy to check that in this limit and shifting the radial
coordinate $r=\hat{\rho}+m-|{\cal M}|$ one recovers the metric
Eqs.~(\ref{eq:ptmet1},\ref{eq:ptmet2},\ref{eq:ptmet3},\ref{eq:ptmet4})
with $|{\cal M}|^{2}-|\Upsilon|^{2}=0$.  Therefore, the supersymmetric
limit of this metric Eqs.~(\ref{eq:ptmet1})-(\ref{eq:ptmetG4}) is a
particular case of the SWIP metric with $1/2$ of the supersymmetries
unbroken.

If we compare with the axion/dilaton IWP solutions presented in
Ref.~\cite{kn:KKOT}, we observe that $1/2$ of the supersymmetries were
also always unbroken.  This meant again that the terms proportional to
the difference $|{\cal M}|^{2}-|\Upsilon|^{2}$ in the metric were always
absent.  The presence of these terms in the solutions that we are going
to study, which implies the breaking of an additional $1/4$ of the
supersymmetries, proves crucial for the existence of regular horizons in
the static cases.  It is also easy to see that the complex scalar
$\lambda$ is also regular on the horizon when only $1/4$ of the
supersymmetries are unbroken: if $|{\cal M}|\neq|\Upsilon|$, then ${\cal
M}\neq \Upsilon$ and the constant in the denominator of

\begin{equation}
\lambda  =  \frac{\lambda_{0} r + \lambda_{0} {\cal M}
+\overline{\lambda_{0}} \Upsilon}{r + {\cal M} + \Upsilon}\, ,
\end{equation}

\noindent never vanishes. Then, on the would-be horizon (which we expect
to be generically placed at $r=0$), $\lambda$ takes the finite value

\begin{equation}
\lambda_{\rm horizon}  =  \frac{\lambda_{0} {\cal M}
+\overline{\lambda_{0}} \Upsilon}{{\cal M} + \Upsilon}\, .
\end{equation}

As we will see, in the rotating case, the additional $1/4$ of broken
supersymmetries does not help in getting regular horizons, though.

From all this discussion we conclude that the most general solution in
the SWIP class describing a point-like object is an asymptotically
Taub-NUT metric with angular momentum.  In general we expect to have at
least $1/4$ of the supersymmetries unbroken and a general proof will be
given later in Section~\ref{sec-susy} with an explicit calculation of
the Killing spinors. This is the main difference with previously known
solutions. (We will compare with rotating asymptotically flat solutions
in Section~\ref{sec-Rotating} )

Instead of studying the most general case, we will study separately two
important particular cases: the non-rotating axion/dilaton Taub-NUT
solution ($\alpha=0\, , l\neq 0$) and the rotating asymptotically flat
solution ($\alpha\neq0\, , l\neq 0$).


\subsection{Extreme axion/dilaton Taub-NUT solution}
\label{sec-Taub-NUT}

When the angular momentum $J=m\alpha$ vanishes (but the mass $m$ does
not vanish), using the coordinate $\rho$ is more adequate, and the
solution takes the form:

\begin{eqnarray}
ds^{2} & = &
\left(1 +\frac{2m}{\rho}
+\frac{|{\cal M}|^{2}-|\Upsilon|^{2}}{\rho^{2}} \right)^{-1}
(dt +2l\cos\theta d\varphi)^{2} \nonumber \\
& & \nonumber \\
& &-\left(1 +\frac{2m}{\rho}
+\frac{|{\cal M}|^{2}-|\Upsilon|^{2}}{\rho^{2}} \right)
(d\rho^{2}+\rho^{2}d\Omega^{2})\, , \\
& & \nonumber  \\
\lambda & = &  \frac{\lambda_{0} r + \lambda_{0} {\cal M}
+\overline{\lambda_{0}} \Upsilon}{r + {\cal M} + \Upsilon}\, ,
\end{eqnarray}

The expressions for the $A^{(n)}_{t}$'s and $\tilde{A}^{(n)}_{t}$'s
are quite complicated and can be readily obtained from the general
solution.

This is the most general extreme axion/dilaton Taub-NUT solution
\cite{kn:NUT} obtained so far.  If we compare the metric with the metric
of the extreme Taub-NUT solution in Eq.~(14) of Ref.~\cite{kn:KKOT} we
can immediately see that the difference is the additional $(|{\cal
M}|^{2}-|\Upsilon|^{2})/\rho^{2}$ term in $e^{-2U}$.  This term vanishes
when both central charge eigenvalues are equal and $1/2$ of the
supersymmetries are unbroken.  This is always the case when there is
only one vector field, as in Refs.~\cite{kn:KKOT,kn:GK} etc.  The effect
of this additional term is dramatic: when it is absent, the coordinate
singularity at $\rho=0$ is also a curvature singularity.  One can easily
see that the area of the two-spheres of constant $t$ and $\rho$ is
$4\pi\rho(\rho+2m)$ and it goes to zero when $\rho$ goes to zero.  If
the additional term that breaks an additional $1/4$ of the
supersymmetries is present, in the limit $\rho=0$ one finds a finite
area:

\begin{equation}
A=4\pi(|{\cal M}|^{2}-|\Upsilon|^{2})\, .
\end{equation}

In fact, it is easy to rewrite the area formula in terms of the
central-charge eigenvalues

\begin{equation}
\label{eq:area}
A = 4\pi ||{\cal Z}_{1}|^{2} -|{\cal Z}_{2}|^{2}|\, ,
\end{equation}

\noindent making evident that, when there is $1/2$ of unbroken
supersymmetry ($|{\cal Z}_{1}|=|{\cal Z}_{2}|$), and only then, the
area vanishes\footnote{We stress again that this is a property of {\it
pure} $N=4$ supergravity that disppears when there is matter.}.

On the other hand, with only one vector field one cannot set to zero the
scalar charge, because it is equal to ${\cal M}$. Only with at least two
vector fields one can set $\Upsilon=0$ and recover solutions of $N=2$
supergravity (all in the IWP class), in particular the charged NUT
metric of Ref.~\cite{kn:Br}.

When $l\neq 0$ this metric does not admit a black hole interpretation
since it has additional naked singularities along the axes
$\theta=0,\pi$ which can be removed using Misner's procedure \cite{kn:M}
at the expense of changing the asymptotics and introducing closed
time-like curves.

When $l=0$ this is the area of the extreme black hole horizon
\cite{kn:KLOPP,kn:KO}.  When only $1/4$ of the supersymmetries are
unbroken and $|{\cal M}|\neq |\Upsilon|$, this area is finite, the
scalars are regular on this surface and $\rho=0$ does not correspond to a
point.  An infinite throat of finite section appears in this limit.

Finally, we observe that the area formula can be written in terms of the
conserved charges\footnote{The conserved charges are defined by
\begin{equation}
\begin{array}{rcl}
{}^{\star}F^{(n)}_{tr} & \sim  & i\frac{\tilde{p}^{(n)}}{r^{2}}\, , \\
& & \\
\left( e^{-2\phi} F -ia{}^{\star}F \right)^{(n)}_{tr} & \sim & 
\frac{\tilde{q}^{(n)}}{r^{2}}\, , \\
\end{array}
\end{equation}
and with them we build the charge vectors
\begin{equation}
\vec{\tilde{q}} = 
\left( 
\begin{array}{c}
\tilde{q}^{(1)} \\
\vdots \\
\tilde{q}^{(N)} \\
\end{array}
\right)\, ,
\hspace{1cm}
\vec{\tilde{p}} = 
\left( 
\begin{array}{c}
\tilde{p}^{(1)} \\
\vdots \\
\tilde{p}^{(N)} \\
\end{array}
\right)\, .
\end{equation}
} 
$\tilde{q}^{(n)},\tilde{p}^{(n)}$ as follows:

\begin{equation}
A = 8\pi \sqrt{\left( \vec{\tilde{q}} \cdot \vec{\tilde{q}} \right) 
\left( \vec{\tilde{p}} \cdot \vec{\tilde{p}} \right) 
-\left(\vec{\tilde{q}} \cdot \vec{\tilde{p}} \right)^{2}}\, .
\end{equation}

In this formula, the independence of the area of the horizon of the
string coupling constant and the moduli is evident, but the manifest
duality invariance seems to be lost\footnote{The invariance under
duality of Eq.~(\ref{eq:area}) is manifest because duality
transformations only permute the absolute value of the central-charge
eigenvalues.}. It is, though, not too complicated to rewrite yet again
the area formula in a manifestly duality-invariant and
moduli-independent fashion:

\begin{equation}
A = 8\pi \sqrt{{\rm det} \left[ 
\left( 
\begin{array}{c}
\vec{\tilde{p}}^{\hspace{2pt}t} \\
\vec{\tilde{q}}^{\hspace{2pt}t} \\
\end{array}
\right) 
\left( 
\vec{\tilde{p}} \,\,
\vec{\tilde{q}} 
\right)
\right]}\, , 
\end{equation}

\noindent where the action of the duality group on the 
charge vector $\left(\begin{array}{c}\vec{\tilde{p}} \\
\vec{\tilde{q}} \\ \end{array} \right)$ is given by

\begin{equation}
\left(
\begin{array}{c}
\vec{\tilde{p}} \\ 
\vec{\tilde{q}} \\ 
\end{array} 
\right)^{\prime}
=
R\otimes S
\left(
\begin{array}{c}
\vec{\tilde{p}} \\ 
\vec{\tilde{q}} \\ 
\end{array} 
\right)\, ,
\end{equation}

\noindent where $R$ is an $SO(6)$ rotation matrix and $S$ is an
$SL(2,\R)$ (unimodular) matrix.


\subsection{Rotating asymptotically flat solution}
\label{sec-Rotating}

When the NUT charge vanishes ($l=0$) in the general solution we get

\begin{equation}
ds^{2} = \frac{\Delta -\alpha^{2}\sin^{2}\theta}{\Sigma}
\left( dt -\omega d\varphi \right)^{2} -
\Sigma \left(\frac{d\hat{\rho}^{2}}{\Delta} +d\theta^{2} +
\frac{\Delta\sin^{2}\theta d\varphi^{2}}{\Delta
-\alpha^{2}\sin^{2}\theta}\right)\, ,
\label{eq:ptmetR1}
\end{equation}

\noindent where

\begin{eqnarray}
\omega & = & -\omega_{\varphi} = \frac{2\alpha\sin^{2}\theta}{\Delta
-\alpha^{2}\sin^{2}\theta}
\left\{m [\hat{\rho} -(m + |\Upsilon|) ]
+{\textstyle\frac{1}{2}}(m^{2} -|\Upsilon|^{2}) \right\}\, ,
\label{eq:ptmetR2} \\
& & \nonumber \\
\Delta & = & \hat{\rho}[\hat{\rho} -2(m+|\Upsilon|)] +\alpha^{2}
+(m+|\Upsilon|)^{2}\, ,
\label{eq:ptmetR3}
\\
& & \nonumber \\
\Sigma & = & \hat{\rho} (\hat{\rho} -2 |\Upsilon|)
+\alpha^{2}\cos^{2}\theta\, .
\label{eq:ptmetR4}
\end{eqnarray}

Again, the expressions for the potentials $A^{(n)}_{t}$ and
$\tilde{A}^{(n)}_{t}$ are complicated and we refer the reader to the
general expression.

When $\alpha=0$, we recover the same general class of static black holes
as in the previous section for $l=0$.  When $m=|\Upsilon|$ this metric
is essentially the one in Eq.~(31) of Ref.~\cite{kn:KKOT} which has
naked singularities.  In this limit we also recover the metric of the
solution in Eqs.~(3.11)-(3.16) of Ref.~\cite{kn:HS} (after going to the
string frame).  Other rotating solutions of the low-energy heterotic
string effective action \cite{kn:S,kn:CY}, after truncation (so they can
be considered solutions of pure $N=4,d=4$) seem to give the same metric
in the supersymmetric limit, breaking only $1/2$ of the supersymmetries.
The exception is the supersymmetric limit of the general rotating
solution in Ref.~\cite{kn:JMP} (see also \cite{kn:S2}), but the
situation is unclear because the metric was not explicitly written down
in the supersymmetric limit.

The new rotating solutions in this class are, therefore, those with
$m^{2}-|\Upsilon|^{2}\neq 0$.

%

First of all, we see that, for $\Upsilon=0$ (that is, constant scalar
$\lambda$) one recovers the Kerr-Newman metric with $m=|q|$.  This was
expected since, as we pointed out in Section~\ref{sec-iwp}, the usual
IWP metrics (embedded in $N=4$ supergravity) are obtained when ${\cal
  H}_{1}=i{\cal H}_{2}=\frac{1}{\sqrt{2}}V^{-1}$. This metric has a
naked ring singularity\footnote{The fact that it is a ring, and not
  just a point can be seen by a further shift of the radial
  coordinate.} at $r=\cos\theta=0$. On the other hand, it has
$m^2-|\Upsilon|^{2}\neq 0$, which, according to our central charge
analysis at the beginning of this Section should mean that it has only
$1/4$ of the supersymmetries unbroken when embedded in $N=4$
supergravity. We will give a direct proof of this in the next Section.

When $\Upsilon\neq 0$ the situation becomes even worse: there are two
naked singularities at

\begin{equation}
\hat{\rho} =|\Upsilon| \pm \sqrt{|\Upsilon|^{2} -\alpha^{2}\cos^{2}\theta}\,
.
\end{equation}

These singularities become one ring-shaped singularity
($\hat{\rho}=0,\theta=\pi/2$) when $\Upsilon=0$, but for general
$\Upsilon$, the range of values of $\theta$ and $\hat{\rho}$ such that
$g_{tt}$ and $g_{\varphi\varphi}$ diverge is bigger and the
singularities are in open surfaces satisfying the above equation.  The
surfaces are closed when $\theta$ can take all values from $0$ to $\pi$,
that is, when $|\Upsilon| \geq |\alpha|$.

All rotating, supersymmetric, point-like objects in this class of
metrics seem to have naked singularities.  A similar result was recently
obtained in the framework of the low-energy heterotic string effective
action in Ref.~\cite{kn:CY} and previously in a more restricted case in
Ref.~\cite{kn:S2} and Ref.~\cite{kn:HS} where the difference with the
situation in higher dimensions was also discussed.


\section{Supersymmetry}
\label{sec-susy}

In this section we study the unbroken supersymmetries of the SWIP
solutions.  We first consider only two vector fields.  Pure $N=4,d=4$
supergravity \cite{kn:CSF} has six vector fields that, in the
supersymmetry transformation laws appear in two different fashions:
three of them are associated to the three metrics $\alpha^{a}_{IJ}$
and the other three are associated to the three matrices
$\beta^{a}_{IJ}$ given in Ref.~\cite{kn:GSO}.  Our choice
\cite{kn:KLOPP} is to identify the first vector field with the vector
field that couples to $\alpha_{IJ}^{3}\equiv \alpha_{IJ}$ and the
second vector field with the one that couples to $\beta^{3}_{IJ}\equiv
\beta_{IJ}$.  The corresponding supersymmetry rules are

\begin{eqnarray}
{\textstyle\frac{1}{2}}\delta \Psi_{\mu I} &=& \partial_\mu\epsilon_I
-{\textstyle\frac{1}{2}}\omega_\mu^{+ab}\sigma_{ab}\epsilon_I
-{\textstyle\frac{i}{4}}e^{2\phi}\bigl (\partial_\mu a\bigr )\epsilon_I
\nonumber\\
& & \nonumber \\
&&-{\textstyle\frac{1}{2\sqrt{2}}}e^{-\phi}\sigma^{ab}\biggl (
F_{ab}^{+(1)}\alpha_{IJ} + i F_{ab}^{+(2)}\beta_{IJ}\biggr )
\gamma_\mu\epsilon^J\, ,\\
& & \nonumber \\
{\textstyle\frac{1}{2}}\delta\Lambda_I &=& - {\textstyle\frac{i}{2}}
e^{2\phi}\gamma^\mu\bigl (\partial_\mu\lambda\bigr )\epsilon_I +
\textstyle\frac{1}{\sqrt{2}} e^{-\phi}\sigma^{ab}\biggl (
F_{ab}^{-(1)}\alpha_{IJ} + i F_{ab}^{-(2)}\beta_{IJ}\biggr )\epsilon^J\, .
\nonumber
\end{eqnarray}

Making the obvious choice of vierbein one-forms and vectors basis

\begin{equation}
\left\{
\begin{array}{rcl}
e^{0} & = & e^{\phi}(dt+\omega_{\underline{i}}
dx^{\underline{i}})\, ,\\
& &  \\
e^{i} & = & e^{-\phi} dx^{\underline{i}}\, , \\
\end{array}
\right.
\hspace{1cm}
\left\{
\begin{array}{rcl}
e_{0} & = & e^{-\phi}\partial_{\underline{0}}\, , \\
& &  \\
e_{i} & = & e^{\phi}(-\omega_{\underline{i}}\partial_{\underline{0}}
+\partial_{\underline{i}})\, ,\\
\end{array}
\right.
\end{equation}

\noindent the components of the spin-connection one-form are given by

\begin{eqnarray}
\omega^{+0i} &=& {\textstyle\frac{i}{4}} e^{3U}\biggl [
\partial_{\underline i}\overline V e^0 + i\epsilon_{ijk}\partial_{\underline j}
\overline V e^k \biggr ]\, ,
\nonumber\\
& & \nonumber \\
\omega^{+ij} &=& {\textstyle\frac{1}{4}} e^{3U}\biggl [
\epsilon_{ijk}\partial_{\underline k}\overline V e^0 + 2i \bigl (\partial_{[
{\underline i}}\overline V \bigr ) \delta_{j]k}e^k \biggr ]\, ,
\end{eqnarray}

\noindent where $V= b + i e^{-2U}$. The curvatures for the vector fields
are given by

\begin{eqnarray}
\label{curv}
F_{0i}^{+(n)} &=& -{\textstyle\frac{i}{2}}e^{2\phi} B^{(n)}_{\underline i}\, ,
\nonumber\\
& & \nonumber \\
F_{ij}^{+(n)} &=& {\textstyle\frac{1}{2}} \epsilon_{ijk}e^{2\phi}
B^{(n)}_{\underline k}\, ,
\end{eqnarray}

\noindent with

\begin{equation}
B^{(n)}_{\underline i} = {e^{2U}\over {\overline{\cal H}_2}}\biggl [
\partial_{\underline i}{\overline {\cal H}_2}\bigl ( k^{(n)}{\cal H}_1
+ {\overline k}^{(n)}{\overline {\cal H}_1}\bigr ) -
\partial_{\underline i}{\overline {\cal H}_1}\bigl ( k^{(n)}{\cal H}_2
+ {\overline k}^{(n)}{\overline {\cal H}_2}\bigr )\biggr ]\, .
\end{equation}

We first consider the supersymmetry variation of the gravitino.  The
variation of the time component leads to (assuming that the Killing
spinor is time--independent)

\begin{eqnarray}
{\textstyle\frac{1}{2}}\delta \Psi_{{\underline 0}I} & = &
-{\textstyle\frac{1}{2}}
e^U\omega_0^{+ab}\sigma_{ab}\epsilon_I \nonumber \\
& & \nonumber \\
& &
-{\textstyle\frac{1}{4}}\sqrt 2 e^{U-\phi}\sigma^{ab}
\biggl ( F_{ab}^{+(1)}\alpha_{IJ} + iF_{ab}^{+(2)}\beta_{IJ}\biggr )
\gamma_0\epsilon^J \nonumber \\
& & \nonumber \\
& = & 0\, ,
\end{eqnarray}

\noindent or

\begin{equation}
\label{id1}
\sqrt 2\biggl ( B_{\underline i}^{(1)}\alpha_{IJ} + i B_{\underline i}^{(2)}
\beta_{IJ}\biggr )\epsilon^J = e^{-\phi + 3U}
\bigl (\partial_{\underline i}\overline V\bigr )\gamma^0\epsilon_I\, .
\end{equation}

On the other hand, the variation of the space--components leads
to

\begin{eqnarray}
{\textstyle\frac{1}{2}}\bigl (\delta\Psi_{\underline i} -
\omega_{\underline i}
\delta\Psi_{\underline 0}\bigr ) &=& \partial_{\underline i}\epsilon_I
- {\textstyle\frac{i}{4}}e^{2\phi}\bigl (\partial_{\underline i} a
\bigr )\epsilon_I
+ \gamma^j\biggl [ - e^{-U}\omega_i^{+0j}\gamma^0\epsilon_I
\nonumber\\
& & \nonumber \\
&&+{\textstyle\frac{1}{2}}\sqrt 2 e^{-U-\phi}\bigl (
F_{0j}^{+(1)}\alpha_{IJ} + i F_{0j}^{+(2)}\beta_{IJ}\bigr )
\gamma^i\gamma^0\epsilon^J\biggr ] \nonumber \\
& & \nonumber \\
& = & 0\, .
\end{eqnarray}

Applying the identity

\begin{equation}
\gamma^j\gamma^i F_{0j}^{+(n)} = F_{0i}^{+(n)} - \gamma^j F_{ij}^{+(n)}
\gamma^0\gamma_5\, ,
\end{equation}

\noindent and using the explicit form of $\omega$ and $F$ the
requirement that

\begin{equation}
{\textstyle\frac{1}{2}}\bigl (\delta\Psi_{\underline i} -
\omega_{\underline i} \delta\Psi_{\underline 0}\bigr ) = 0\, ,
\end{equation}

\noindent leads to the following equation for the Killing spinor

\begin{equation}
\partial_{\underline i}\epsilon_I - {\textstyle\frac{i}{4}} e^{2\phi}
\bigl (\partial_{\underline i} a\bigr )\epsilon_I
+ {\textstyle\frac{1}{2}}\sqrt 2 e^{-U-\phi} \biggl (
F_{0i}^{+(1)}\alpha_{IJ} + i F_{0i}^{+(2)}\beta_{IJ}\biggr )
\gamma^0\epsilon^J = 0\, .
\end{equation}

\noindent Substituting the expression (\ref{curv}) and applying
(\ref{id1}) this Killing spinor equation can be rewritten as follows

\begin{equation}
\partial_{\underline i}\epsilon_I -{\textstyle\frac{i}{4}}e^{2\phi}
\bigl (\partial_{\underline i}a\bigr )\epsilon_I +
{\textstyle\frac{i}{4}}e^{2 U}
\bigl (\partial_{\underline i} \overline V \bigr )\epsilon_I  = 0\, .
\end{equation}

\noindent Next, we apply the identity

\begin{equation}
e^{2\phi}\bigl (\partial_{\underline i} a\bigr ) -
e^{2 U}\bigl (\partial_{\underline i} b\bigr ) =
{i\over {\overline {\cal H}_2}} \partial_{\underline i}
{\overline {\cal H}_2} -{i\over {\cal H}_2} \partial_{\underline i}
{\cal H}_2\, ,
\end{equation}

\noindent and find that

\begin{equation}
\partial_{\underline i}\epsilon_I +{\textstyle\frac{1}{4}}
\biggl (
{\partial_{\underline i}{\overline {\cal H}_2}\over
{\overline {\cal H}_2}}
-{\partial_{\underline i}{\cal H}_2\over {\cal H}_2}
\biggr )
\epsilon_I -{\textstyle\frac{1}{2}}
\left( \partial_{\underline i}U\right)\epsilon_I
=0\, .
\end{equation}

Finally, we can solve this equation for the Killing spinor as follows:

\begin{equation}
\epsilon_I = \left( {{\overline {\cal H}_2}\over {\cal H}_2}
\right)^{-1/4} e^{U/2}\ \epsilon_{I(0)}\, ,
\end{equation}

\noindent for constant $\epsilon_{I(0)}$.  We note that the first factor
in the expression for $\epsilon_I$ containing the harmonic function is
exactly what one would expect for a $SL(2,\R)$ covariant spinor
\cite{kn:O2}

The constant spinors $\epsilon_{I(0)}$ satisfy certain algebraic
conditions which are determined by the vanishing of the supersymmetry
rules of the time--com\-po\-nent of the gravitino (see Eq.~(\ref{id1}))
and of $\Lambda_{I}$.  For clarity, we repeat here Eq.~(\ref{id1}), and
give the equation that follows from the vanishing of
$\delta\Lambda_{I}$:

\begin{eqnarray}
\sqrt 2\biggl ( B_{\underline i}^{(1)}\alpha_{IJ} + i B_{\underline
i}^{(2)} \beta_{IJ}\biggr )\epsilon^J &=& e^{-\phi + 3U}
\bigl (\partial_{\underline i}\overline V\bigr )\gamma^0\epsilon_I\, ,
\nonumber\\
& & \label{id2} \\
\sqrt 2 \biggl ( {\overline B^{(1)}_{\underline i}}\alpha_{IJ}
+ i {\overline B^{(2)}_{\underline i}}\beta_{IJ}\biggr )\epsilon^J
&=& e^{\phi + U}\bigl (\partial_{\underline i}\lambda\bigr ) \gamma^0
\epsilon_I\, .\nonumber
\end{eqnarray}

\noindent In terms of the harmonic functions these two equations read as
follows\footnote{To derive this equation we must make a choice of
  convention for the branch cuts of the square root of a complex
  number.  In the specific calculation below we have made use of the
  identity $\overline {{\cal H}}_2/\overline {{\cal H}}_2^{1/2}=
  \overline {{\cal H}}_2^{1/2}$.  The effect of taking another sign at
  the r.h.s. of this equation is that in the final answer for the
  Killing spinors $\gamma^0 \rightarrow - \gamma^0$.}:

\begin{eqnarray}
\bigl [ k^{(1)}{\cal A}_{\underline i} + {\overline k^{(1)}}\overline
{{\cal B}}_{\underline i}\bigr ] \alpha_{IJ}\epsilon^J_{(0)} + i
\bigl [ k^{(2)} {\cal A}_{\underline i} +
\overline {k}^{(2)}\overline {{\cal B}}_{\underline i}\bigr ]\beta_{IJ}
\epsilon_{(0)}^J &=& - {\cal A}_{\underline i}\gamma^0
\epsilon_{I(0)}\, ,\nonumber\\
& & \label{harmeq} \\
\bigl [ \overline{k}^{(1)}\overline{{\cal A}}_{\underline i} + k^{(1)}
{\cal B}_{\underline i}\bigr ] \alpha_{IJ}\epsilon^J_{(0)} + i
\bigl [ \overline{k}^{(2)} \overline{{\cal A}}_{\underline i} +
k^{(2)}{\cal B}_{\underline i}\bigr ]\beta_{IJ}
\epsilon_{(0)}^J &=& -{\cal B}_{\underline i}\gamma^0
\epsilon_{I(0)}\, , \nonumber
\end{eqnarray}

\noindent with

\begin{eqnarray}
{\cal A}_{\underline i} &=& {\cal H}_1\partial_{\underline i}
\overline{{\cal H}}_2 - {\cal H}_2\partial_{\underline i}
\overline{{\cal H}}_1\, ,\nonumber\\
& & \\
{\cal B}_{\underline i} &=& {\cal H}_1\partial_{\underline i}
{\cal H}_2 - {\cal H}_2\partial_{\underline i}
{\cal H}_1\, .\nonumber
\end{eqnarray}

Let us now define

\begin{equation}
{\cal C}_{\underline{i}}\equiv {\cal A}_{\underline{i}} -e^{2i\delta}
\overline{\cal B}_{i}\, ,
\end{equation}

\noindent for some real and non-yet specified parameter $\delta$.
Next, we solve the two constrained complex parameters $k^{(1)}$ and
$k^{(2)}$ in terms of a single real parameter $\gamma$ as follows:

\begin{equation}
k^{(1)} = {\textstyle\frac{1}{2}} e^{i\gamma}\, ,\hskip 1.5truecm
k^{(2)} = {\textstyle\frac{i}{2}} e^{i\gamma}\, .
\end{equation}

Using this parametrization of $k^{(1)}$ and $k^{(2)}$, it is easy
to see that Eqs.~(\ref{harmeq}) reduce to

\begin{eqnarray}
{\cal A}_{\underline{i}}\ \xi_{I} -{\cal C}_{\underline{i}}\ \eta_{I}
& = & 0\, ,\\
& & \nonumber \\
\overline{\cal A}_{\underline{i}}\ \zeta_{I}
-\overline{\cal C}_{\underline{i}}\ \eta_{I}
& = & 0\, ,
\end{eqnarray}

\noindent where  $\xi_{I},\eta_{I}$ and $\zeta_{I}$ are constant spinors
defined as follows:

\begin{eqnarray}
\eta_{I} & = & {\textstyle\frac{1}{2}} e^{-i(\gamma +\delta)}
(\alpha_{IJ} +\beta_{IJ})  \epsilon^{J}_{(0)}\, ,\\
& & \nonumber \\
\zeta_{I} & = & {\textstyle\frac{1}{2}} e^{i(\gamma +\delta)}
(\alpha_{IJ} -\beta_{IJ})  \epsilon^{J}_{(0)}
+e^{i\delta} \gamma^{0} \epsilon_{I(0)}\, .\\
& & \nonumber \\
\xi_{I} & = & \left[\cos (\gamma +\delta) \alpha_{IJ}
-i\sin (\gamma +\delta) \beta_{IJ} \right] \epsilon^{J}_{(0)}
+ e^{i\delta} \gamma^{0} \epsilon_{I(0)}\\
& & \nonumber \\
& = & \eta_{I} +\zeta_{I}\, .
\end{eqnarray}

We have to consider several different cases:

\begin{enumerate}

\item If both ${\cal A}_{\underline{i}}$  and ${\cal C}_{\underline{i}}$
are different from zero for any possible value of $\delta$ and they are
also different for any value of $\delta$, then, since the spinors
$\xi_{I},\eta_{I},\zeta_{I}$ are constant, they have to vanish. Given
that $\xi_{I} = \eta_{I} +\zeta_{I}$, this gives only two independent
conditions:  $\eta_{I}=\zeta_{I}=0$ on the constant spinors
$\epsilon_{I(0)}$.

By making use of the explicit form of the matrices $\alpha_{IJ}$ and
$\beta_{IJ}$ (see \cite{kn:KLOPP}), we can solve these two equations and
we find that the constant parts of the Killing spinors are

\begin{equation}
\epsilon_{(0)}^3 = \epsilon_{(0)}^4 = 0\, ,
\hspace{1cm}
\epsilon_{1(0)} = e^{i\gamma}\gamma^0\epsilon_{(0)}^2\, .
\label{eq:Kill}
\end{equation}

This is the generic case and we just have proven that only $1/4$
of the supersymmetries are unbroken in this case.

\item If both ${\cal A}_{\underline{i}}$ and ${\cal C}_{\underline{i}}$
are equal to each other and different from zero for some value of
$\delta$, then, using that $\xi_{I} = \eta_{I} +\zeta_{I}$ we get again
the same equations and the same amount of unbroken supersymmetry.

\item If ${\cal A}_{\underline{i}}$ vanishes we are again in the same
case.

\item If ${\cal C}_{\underline{i}}$ vanishes for some value of $\delta$,
then, we only get one condition on the constant spinors
$\epsilon_{I(0)}$: $\xi=0$.  In this case, and only in this case, $1/2$
of the supersymmetries are unbroken.

\end{enumerate}

This proofs that we always have {\it at least $1/4$ of the
supersymmetries are unbroken and all the SWIP metrics admit Killing
spinors}.

We can now study how different particular cases fit into this scheme.
We can treat first the case of a single point-like object because we
know in which cases one or two Bogomol'nyi bounds are saturated.  For
one bound saturated we expect two algebraic constraints on the Killing
spinors and for two bounds saturated we expect only one ($1/4$ and $1/2$
of the supersymmetries unbroken respectively).  Substituting the
expressions Eqs.~(\ref{eq:h1point}) and (\ref{eq:h2point}) of ${\cal
H}_1$ and ${\cal H}_2$ in Eqs.~(\ref{eq:h1point}) and (\ref{eq:h2point})
which are expressed in terms of the charges we find that

\begin{eqnarray}
{\cal A}_{\underline i} &=& i\biggl \{\overline {{\cal M}}\
\partial_{\underline i} {1\over \overline {r}} + \bigl (|{\cal M}|^2
- |\Upsilon|^2\bigr )\ {1\over r}\partial_{\underline i}  {1\over
\overline {r}} \biggr\}\, ,\nonumber\\
& &  \\
{\cal B}_{\underline i} &=& i \Upsilon\
\partial_{\underline i}{1\over r}\, .\nonumber
\end{eqnarray}
It is easy to see that when $|{\cal M}|^{2} -|\Upsilon|^{2}=0$, then
${\cal M}=e^{i2\delta}\overline{\Upsilon}$ and we are in the case in
which ${\cal C}_{\underline{i}}=0$ and $1/2$ of the supersymmetries are
unbroken. $|{\cal M}|^{2} -|\Upsilon|^{2}\neq 0$ is the generic case and
only $1/4$ of the supersymmetries are unbroken, as expected form the
Bogomol'nyi bounds analysis.

Another particular case is when ${\cal H}_{1} =e^{i\sigma} {\cal H}_{2}$
for some real constant $\sigma$. Then ${\cal B}_{\underline{i}}=0$ and
${\cal C}_{\underline{i}}={\cal A}_{\underline{i}}$ and only $1/4$ of
the supersymmetries are unbroken. In the $N=2,d=4$ IWP solutions

\begin{equation}
{\cal H}_{1}=i{\cal H}_{2}={\textstyle\frac{1}{\sqrt{2}}} V^{-1}\, ,
\end{equation}

\noindent that is $\sigma=\pi/2$, and, as expected, they have only $1/4$
of the supersymmetries unbroken when embedded in $N=4,d=4$ supergravity.


\section{Conclusion}
\label{sec-conclusion}

We have presented a family of supersymmetric stationary solutions that
generalizes to $N=4$ supergravity the IWP solutions of $N=2$
supergravity in the sense that they are the most general solutions of
its kind and that they are manifestly invariant (as a family) under
all the duality symmetries of $N=4,d=4$ supergravity.

We have studied the supersymmetry properties of the general solution
and the geometry of the most general point-like solution in this
class, arriving to the conclusion that no rotating supersymmetric
black holes exist in pure $N=4,d=4$ supergravity (see also
\cite{kn:KKOT,kn:S2,kn:CY}). The situation in four dimensions seems to
be radically different from the situation in five dimensions, where
supersymmetric rotating black holes with regular horizon have been
found in Ref.~\cite{kn:HS}. The physical reason for this has not yet
been understood.

We have also argued that the interpretation of our solution for the
two vectors case as a solution of $N=2,d=4$ supergravity coupled to
one vector field and in terms of special geometry provides a most
interesting clue to get the most general supersymmetric stationary
solution of $N=2,d=4$ supergravity coupled to any number of vector
multiplets.

We have not studied other solutions in this class describing extended
objects like strings and membranes. These can be obtained by choosing
complex harmonic functions that depend on less than three spatial
coordinates. These solutions are, of course, also supersymmetric
because the general analysis performed in Section~\ref{sec-susy}
applies to them. They are also manifestly $S$ and $T$~duality
invariant. We have not discussed massless solutions either. Exploiting
to its full extent this class of solutions will require much more
work, but all the results that will eventually be obtained will also
be manifestly duality invariant. We believe that this is really
progress towards a full characterization of the most general
supersymmetric solution in $N=8$ supergravity, which should be
manifestly $U$~duality invariant. Having this solution at hands would
be of immense value as a testing ground for all the current ideas on
duality and the construction of supersymmetric black holes from
$D$~branes.

It would also be interesting to take these solutions out of the
supersymmetric limit.  The non-supersymmetric metrics should also be
manifestly duality invariant and would describe single stationary
black holes, asymptotically Taub-NUT metrics, strings or membranes
because there would be no balance of forces and it would be impossible
to have more than one of such objects in equilibrium.  Many of these
non-supersymmetric solutions are known in the point-like case (black
holes or asymptotically Taub-NUT metrics) (see, for instance,
\cite{kn:KO}). This is, we hope, one of the main goals of obtaining
supersymmetric solutions. They are easier to obtain because of the
additional (super) symmetry that constrains the equations of motion
and they should help in finding the general non-supersymmetric
solutions which are, perhaps, the most interesting from a physical
point of view.

After the completion of this work we discovered that some of the
stationary solutions presented in this work have also been found, in a
different setting, in \cite{kn:T2}, by directly solving the Killing
spinor equations.


\section*{Acknowledgements}

We would like to thank S.~Ferrara and A.W.~Peet for most useful
discussions.  E.B.~and T.O.~would like to express their gratitude to
the Physics Department of Stanford University for the stimulating
working environment and for financial support.  The work of E.B.~has
been made possible by a fellowship of the Royal Netherlands Academy of
Arts and Sciences (KNAW).  The work of R.K.~was supported by the NSF
grant PHY-9219345.  The work of T.O.~and E.B.~has also been partially
supported by a NATO Collaboration Research Grant.  T.O.~would also
like to thank M.M.~Fern\'andez for her support.


\end{document}